\documentclass[journal=acsodf,manuscript=article]{achemso}
\usepackage[version=3]{mhchem}
\usepackage{graphicx,caption,subfigure,lscape}  % needed for figures  
\usepackage{amssymb, ulem}
 \usepackage[outdir=./]{epstopdf}% for math
 \usepackage[usenames,dvipsnames]{color}
 \usepackage[hyperfootnotes=false]{hyperref}
 \usepackage[hyperfootnotes=false]{hyperref}
 \hypersetup{  colorlinks=true,  citecolor=Violet,  linkcolor=Red,  filecolor= green,  urlcolor=Blue}

\author{S. K. Pandey}
\author{Ruma Das}
\author{Priya Mahadevan}
\affiliation
{Department of Condensed Matter Physics and Material Science, \\ S. N. Bose National Center for Basic Sciences,Kolkata}
\email{priya.mahadevan@gmail.com}

\title{Layer dependent electronic structure changes in transition metal dichalcogenides- The microscopic origin}

\begin{document}

\begin{abstract}
We have examined the electronic structure evolution in transition metal dichalcogenides MX$_2$, where M=Mo,W and X=S,Se and Te.
 These are generally referred to as van der Waals heterostructures on the one hand, yet one has band gap
 changes as large as 0.6 eV with thickness in some instances. This does not seem to be consistent with a description where the dominant interactions
 are van der Waals interactions. Mapping onto a tight binding model allows us to quantify the electronic structure changes
 which are found to be dictated solely by interlayer hopping interactions. Different environments that an atom encounters
 could change the Madelung potential and therefore the onsite energies. This could happen while going from monolayer to bilayer as well as
 in cases where the stackings are different from what is found in 2H structures. These effects are quantitatively found to be negligible, enabling us
 to quantify the thickness dependent electronic structure changes as arising from interlayer interactions alone.
\end{abstract}

\section{Introduction}
Although  the transition metal dichalcogenides have been studied for fifty plus years, it is amazing that novel phenomena are still being 
discovered in these materials. Additionally the current thrust for alternate technologies have led to their successful use in catalysis,
in addition to explorations in photovoltaics, nanoelectronics etc \cite{cite3,cite4,cite5,cite6,cite7}.
Analogous to the nanomaterials where one finds a size dependence
of the band gap \cite{nanda,nmat-ithuria, ddsarma},
one finds thickness dependent changes in the electronic structure of the layered transition
metal dichalcogenides\cite{mos2bulk,Tongay,mos2mono, mosemono} . Additionally one finds a thickness dependent band gap which changes character.
The bulk band gap (optical) of MoS$_2$ is found to be an indirect one of 1.3 eV \cite{mos2bulk} which increases to
1.6 eV in the bilayer limit \cite{Tongay}. The nature of the band gap changes and becomes a direct one of 1.9 eV at the
monolayer limit \cite{mos2mono}. The fact that the monolayers of Mo and W-based transition metal dichalcogenides have a direct band gap
(with the exception of WSe$_2$ \cite{wse2_indir}) is evident from the sharp peak that one finds in the photoluminescence spectra\cite{splend} .
MoSe$_2$ also has an indirect band gap of 1.1 eV \cite{mos2bulk} in the bulk limit whereas in the monolayer limit
it has direct band gap of about 1.66 eV \cite{mosemono}. A smaller change is found in the bandgap of MoTe$_2$ in contrast
to MoS$_2$ and MoSe$_2$. Here one finds the indirect bulk band gap of 0.9 eV changes to 1.1 eV at the monolayer limit. 
Considering the W-based analogues, one has a change of 0.75 eV in WS$_2$, while one has a smaller change of 0.45 eV in WSe$_2$
\cite{ws2_mono,wilson,ws2_se2}. 

There could be different types of MX$_2$ (M = Mo, W, Ti etc., X = S, Se, etc.)
sandwiches depending on the coordination of the transition metal atom with 
the chalcogens as well as the stacking of atoms \cite{chhowalla,cesare}.
In this work we focus our attention on the 2H polymorph 
of MX$_2$ (M=Mo,W, X=S,Se,Te)\cite{mos2_se2_exp, wse2_exp, ws2_se2_exp, wte2_mote2} 
 where the symmetry
about the Mo/W site is trigonal prismatic, though certain 
generic features are found to be valid across different
types of stacking~\cite{cesare}. The bonding within each monolayer is strongly covalent.
However, the coupling between layers is
believed to be due to weak van der Waals interaction. This has led to the multilayers
being called van der Waals heterostructures. Hence the changes in the band gap, which was found to be as large as 0.75~eV in
WS$_2$ \cite{ws2_mono} seems surprising.

A popular method to examine the size dependent electronic structure of semiconductor nanostructures has been the tight-
binding model. The size dependence within the model emerges from two effects.
The first is the changed coordination of the atom which could affect the Madelung potential and therefore the
onsite energies of the levels on that atom. The second is the change in the bandwidth due to reduced coordination
faced by some of the atoms. Usually only the latter effects are considered. In the context of the transition
metal dichalcogenides, however, the effects due to the former become important as unlike in the nanostructures
where a small fraction of the atoms have different coordination, here, a significant fraction have a coordination
different from the bulk. A mapping onto a tight binding model allows us to explore for the first time the
role of various contributions leading to quantum confinement.

The issue of the origin of the size dependence of the bandgap in TMDs has been addressed earlier in the literature~\cite{heine}.
Zhang and Zunger~\cite{zhang1} examined the electronic structure as a function
of thickness and identified the variations seen to arise from two factors  -
the kinetic energy controlled quantum confinement as well as the
potential energy controlled level repulsion. Kang and coworkers~\cite{wei}
suggest that the crystal symmetries of hexagonal layered materials
lead to a weak interlayer coupling of the band extrema at the K point.
Modifications in the local symmetry for instance break these symmetries
and allow for the emergence of an electronic structure that depends on
thickness.
Cappelluti  {\it et al.}  \cite{cappe} have fit the $ab-initio$ band structure for a monolayer of MoS$_2$ to a nearest
neighbour tight-binding model. The same parameters were used for the bilayer with additionally interlayer interactions thrown in.
This was then used to infer that it is the interlayer interactions that were 
responsible for the electronic structure changes in MoS$_2$ with thickness, especially the direct to the indirect bandgap transition.
Among other attempts of tight binding based studies, 
there have been a large number of studies determining the relevant tight-binding parameters~\cite{rostami, liu_tb, prb_tb, aip_tb}. 
These studies have started with different basis set choices.
The values of the parameters entering the tight-binding Hamiltonian have been determined either by fitting the $ab-initio$ band structure in a small energy 
window in the vicinity of the band gap or constraining the model to reproduce some physical parameters such as the band gap, hole/electron effective mass, the 
location of the valence band maximum and the conduction band minimum. In some instances, a non-orthogonal basis set has been used\cite{rostami,aip_tb}. This has the
disadvantage that the extracted onsite energies do not correspond to the natural orbital energy and one cannot discuss trends across the series where the
anion is changed for instance. Further, one usually prefers to work with a model using an orthogonal basis while including many body effects.

We build on the ideas given in the literature. As mentioned earlier, the changed environment could lead to modifications in the
Madelung potentials associated with the atoms and consequently changes in the onsite energy. This was not considered earlier.
Further a mapping of the $ab-initio$ band structure over a wider range of energy allows us to describe the trends in the electronic
structure better. For this reason we carry out a mapping of the electronic structure as a function of thickness for Mo- and W- based
TMDs. We have fixed the interlayer distances at the values at which all the
calculations are reported in ths manuscript, 
and then calculated the electronic structure 
after switching off the van der Waals interactions.
A similar mapping of the $ab-initio$ band structure onto a tight-binding model
was carried out. 
We find that there is 
no change in the onsite energies of the bilayers of TMDs from the values that
we had including van der Waals interactions.
This implies that the van der Waals interactions determine the interlayer seperation. However, they do not modify the electronic structure.
Additionally in each of the systems studied, we find that switching off the interlayer hopping interactions for the bilayers and beyond, we are able to
recover the monolayer bandstructure. This demonstrates that even in materials in which the interaction between layers is believed
to be van der Waals type, it is covalent interactions between the layers that determines the evolution of the electronic
structure with an increase in the number of layers. Consequently a knowledge of the tight-binding Hamiltonian
for the monolayer as well as the strength
of the interlayer hopping interactions allows us to construct the Hamiltonian for any number of layers. While the discussion so far has focussed
on the stacking that is favoured in the bulk structures, the 2H stacking, alternate stackings are also possible. Considering
bilayers of MoSe$_2$, steric effects between the electrons on the two layers determine the interlayer separations for different stackings.
However, one finds that the nature of stacking has very small effect on the onsite energies which are again found to be the same
as the monolayer by a similar analysis. These results imply that 
even while modeling twisted bilayers of these materials, the variations 
we find in the electronic structure emerge from the interlayer 
interactions which has been the approach adopted in the literature. 
Our analysis in the present work show that the inclusion of 
weak van der Waals interactions in the Hamiltonian governing 
the electronic structure does not change the onsite energies 
and is only responsible for determining the interlayer separation. 
Building on these arguments, our conclusion that the 
electronic structure evolution 
with number of layers of TMDs are mainly determined by the interlayer hopping interactions stands robust.

\section{Results and Discussion}
\begin{figure}[ht]
\centering
\includegraphics[height=16.cm,width=7.cm,angle=-270.0]{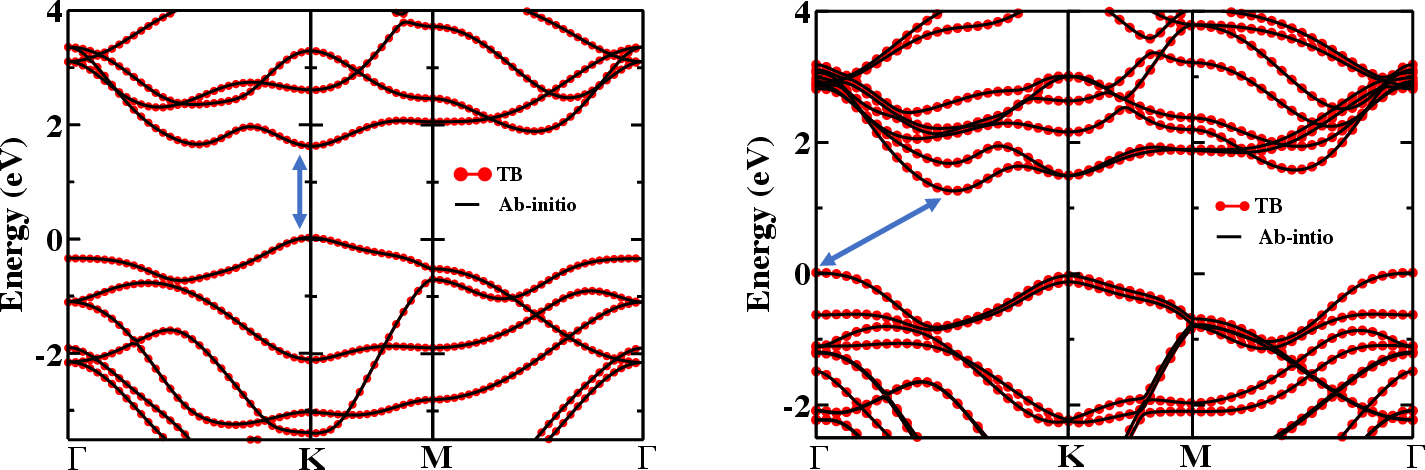}
\caption{ The $ab-initio$ (solid line) and tight binding band structure (red circles) for a) monolayer MoSe$_2$ and b) bilayer  MoSe$_2$. Zero of the energy is the valence band maximum. Arrows represent the VBM and CBM.}
  \label{fig:1}
\end{figure}

The $ab-initio$ band dispersions for monolayer MoSe$_2$ along various
symmetry directions are plotted in Fig. \ref{fig:1}(a). One finds that the
valence band maximum (VBM) and the conduction band minimum (CBM) are
both located at K point and the system is a direct band gap semiconductor.
This is consistent with experiment which also finds the system to be
a direct band gap semiconductor with a gap of 1.66 eV.
While the experimental band gap is the optical band gap\cite{mosemono}., in our calculations
we are calculating the single particle gap.
The present calculations
which use generalized gradient approximation (GGA) for the exchange correlation functional find a gap of
1.59 eV which is close to the experimental value. The agreement is
however fortuitous as one usually has an underestimation of the
band gap due to self-interaction effects among various other
approximations which enter the use of the generalized gradient
approximation in the absence of an exact exchange correlation
functional. In order to quantify the changes in the electronic
structure, we have mapped the $ab-initio$ band structure onto
a tight binding model with Mo $d$ and Se $p$ states in the basis.
The tight binding band structure shown by red line with circles is superposed
on the calculated $ab-initio$ band structure in Fig. \ref{fig:1}(a).
We have a good description of the $ab-initio$ band structure in the
energy window from -3.5 eV to 4 eV. This gives us confidence in the
extracted parameters and allows us to discuss changes in the electronic
structure in terms of these parameters.

There are several ways to construct the bilayer of MoSe$_2$. Each monolayer
can be visualized as a three atomic layer stacking of Mo and Se atoms
where Mo atoms are sandwiched between layers of Se atoms. The Se
atoms generate a trigonal prismatic crystal field at the Mo site. The
stacking that we have considered has the Mo atom in one layer above
that in the layer beneath. However, the Mo-Se motif is rotated by 180$^{\circ}$
in the upper layer with respect to the layer beneath. This is referred to
as the AA$'$ stacking and has been shown to have the lowest energy among
various stacking patterns considered \cite{chhowalla,cesare}.

\begin{figure}[!ht]
\centering
\includegraphics[height=8.cm,width=9.cm]{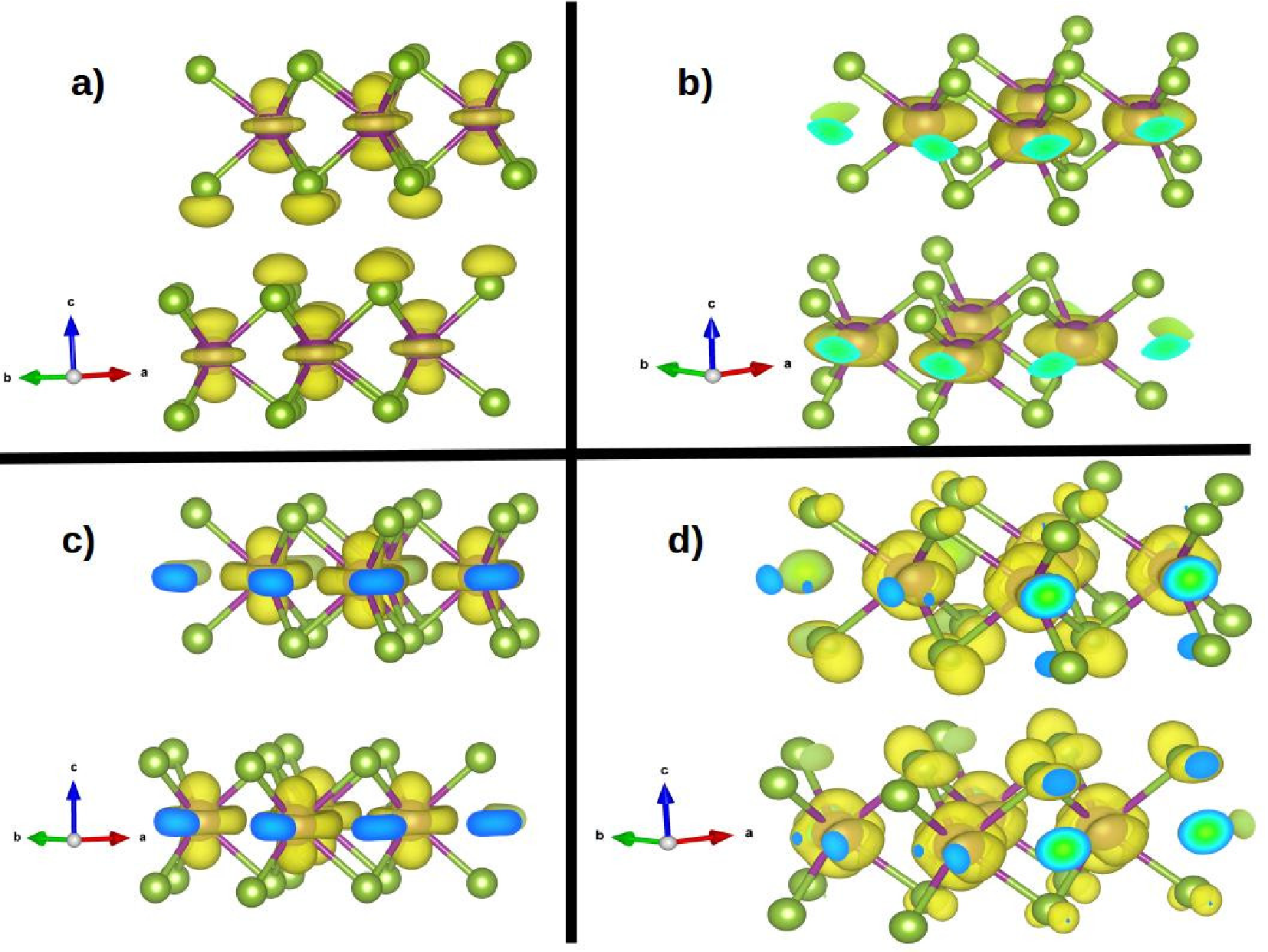}
\caption{ The charge density plots for (a) highest occupied band at G, (b) valence band maximum at K, (c) lowest unoccupied band at T and (d) conduction
  band minimum at K for monolayer MoSe$_2$.}
  \label{fig:3}
\end{figure}

Considering a bilayer of MoSe$_2$, we have calculated the band dispersions
along various symmetry directions. This is shown in Fig. \ref{fig:1}(b). We find that the
VBM which was at  K point has now shifted to $\Gamma$. The CBM is also
shifted to T point which lies along the line from $\Gamma$ to K.
This leads
to an indirect band gap of 1.25 eV in contrast to the experimental value
of 1.55 eV \cite{Tongay}. The changeover in the VBM positions can easily be understood
by examining the character of the states contributing to this point.
This is shown in Fig. \ref{fig:3} where we plot the charge density for the highest
occupied band at $\Gamma$ in panel (a).

These are seen to emerge from the
interactions between the $d_{z^2}$ orbitals on Mo and $p_z$ orbitals on Se.
As these involve orbitals which are directed out of plane, one finds that
these levels in the lower layer interact with the $d_{z^2}$/$p_z$
orbitals in the layer above. As a result, the highest occupied band at $\Gamma$ 
point moves to higher energies relative to that at K point, and consequently the VBM
shifts to $\Gamma$ point when we move from monolayer to bilayer.
The highest occupied band at the K point is contributed by
interactions between in-plane orbitals as is evident from Fig.~\ref{fig:3}(b).
Hence it shows no shift in the bilayer from its position
for the monolayer. A similar analysis of the charge density
contributing to the lowest unoccupied band at T and K symmetry
points is shown in panels (c) and (d) of Fig. \ref{fig:3}. We find that in-plane
orbitals contribute to the lowest unoccupied band at K point while
out-of-plane orbitals contribute to the lowest unoccupied band at T point.
Hence, the increased interaction arising from the presence of the second
layer moves T point relative to K point, making the former the conduction
band minimum. These ideas are consistent with the analysis of Padhila
{\it et al.} \cite{padilha} who examined the movement of various band extrema as a function
of the number of layers in MoS$_2$.

While these ideas are qualitative, we examine the
extracted onsite energies and hopping
interaction strengths in order to make a more quantitative statement
of the role of various effects which determine the
electronic structure. As  mentioned earlier, we have a good
description of the $ab-initio$ electronic structure within the tight binding
model. This gives us confidence in the extracted
parameters.
The onsite energies for Se $p$ as well as Mo $d$ orbitals extracted by the
tight binding mapping for monolayer as well as bilayer MoSe$_2$ are
given in Table \ref{onsite}.
In order to undertand the role of weak van der Waals interaction 
on the electronic structure of MoSe$_2$, we first fixed 
the interlayer seperation at the value used for the
calculations for the bilayer in the manuscript and then switched 
off the van der Waals interactions. 
The extracted energies were found to 
be the same as given in Table~\ref{onsite} but with a constant shift. 
It is hence clear from this analysis that the van der Waals term does not 
modify the Hamiltonian describing the electronic structure of TMDs but enters
only the total energy and plays the role of getting the correct 
interlayer seperation in these systems.

\begin{figure}[ht]
\centering
\includegraphics[height=9.0cm,width=7.0cm,angle=-270.0]{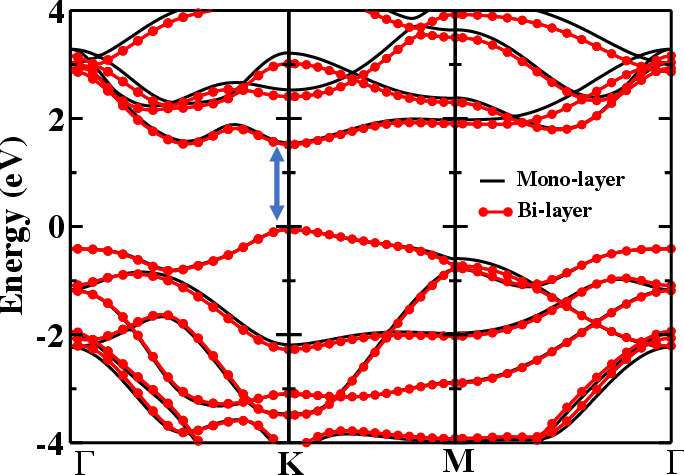}
\caption{The $ab-initio$ band structure (solid line) for monolayer MoSe$_2$ compared with the tight binding band structure (red circles) of bilayer
with interlayer interactions switched off. The zero of energy is the valence band maximum. Arrows represent the VBM and CBM.}
  \label{fig:4}
\end{figure}

After inspection of the onsite energies, one can conclude that
the Madelung potential differences for the atoms in the bilayer
as compared to the monolayer are very small. This validates the 
approximation of using the same onsite energies for each layer in
several earlier studies in the literature. 
In order to examine what is it that
leads to the differences in the electronic structure in going
from monolayer to bilayer, we have considered the tight binding Hamiltonian
for the bilayer. All the interlayer interactions have been switched off in this
Hamiltonian and the ensuing band structure has been plotted along various
symmetry directions in Fig. \ref{fig:4}.
The band structure  for the monolayer has been
superposed for comparison. The two band structures look almost identical
suggesting that the only difference between the electronic structure of
the monolayer and that for the bilayer emerge from interlayer interactions.
The dominant interaction strengths are found
between the $p_z$ orbitals on first (3.72 \AA{}) neighbor 
as well as second (5.92 \AA{}) neighbor Se atoms.

\begin{figure}[ht]
\centering
\includegraphics[height=9.0cm,width=7.0cm,angle=-270.0]{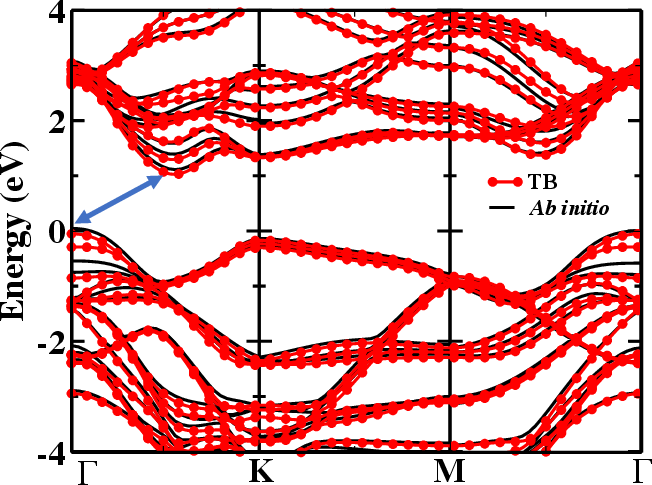}
\caption{ $ab-initio$ (solid line) and tight binding (red circles) band structure for trilayer MoSe$_2$. The tight binding Hamiltonian for the trilayer has been
constructed from monolayer for the layers and interlayer interactions taken from the bilayer. Zero of energy is the valence band maximum. Arrows represent the VBM and CBM.}
  \label{fig:5}
\end{figure}

In order to examine this hypothesis further, we constructed a trilayer
heterostructure of MoSe$_2$. The $ab-initio$ band structure for the
trilayer was calculated along various symmetry directions. This is
shown in Fig.\ref{fig:5}.
For comparison and to examine the hypothesis made
vis-a-vis the origin of the changes in the electronic structure
as each layer is added, we set up the tight binding Hamiltonian for the
trilayer. This was done by considering the Hamiltonian for the monolayer
for each of the layers. The interlayer interactions extracted for the
bilayer were then used to couple the layers. The band structure calculated
within this model was superposed on the $ab-initio$ band structure for
the trilayer in Fig.  \ref{fig:5}. The comparison is reasonably good, justifying our
hypothesis.
These results clearly show that the electronic structure changes
in going from the monolayer to bilayer and beyond are derived from interlayer hopping
interactions alone.

This is a surprising result as the nomenclature used for these systems is van der Waals heterostructures. This would have us believe
that the dominant interaction is van der Waals interactions. However our analysis suggests that covalent interactions determine the modification in the
electronic structure with thickness. This is not entirely surprising as the nearest neighbor separations between Se atoms of two layers is 3.718 \AA{}.
 While the nearest neighbor distance of 
two Se atoms in the same layer is $\simeq$ 3.289 $\AA$, assuming a Harrison-type 
scaling law for the distance, one finds that the hopping interaction strengths
for the interlayer hopping interaction strengths 
drop to 70\% of the values within the layer.
Hence the presence of finite hopping interaction strengths for electrons in the two layers is not entirely surprising. This immediately raises the
question of the role played by van der Waals interactions. This interlayer separation for the bilayer is found to be 3.99 \AA{} when van der Waals
interactions are not included and 3.19 \AA{} when they are. Hence their inclusion merely predicts the interlayer separation.

\begin{figure}[ht]
\centering
\includegraphics[height=10.0cm,width=15.0cm]{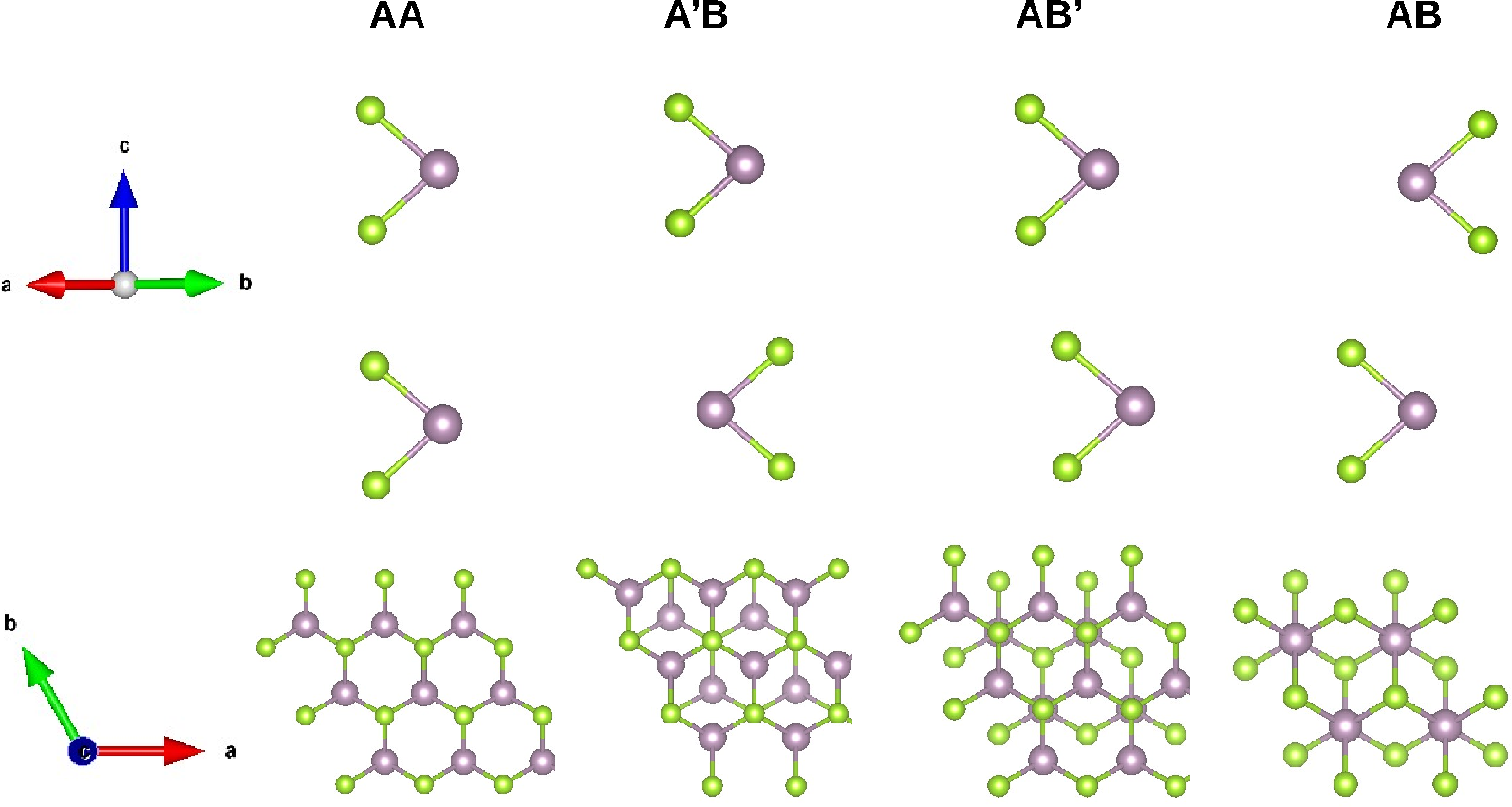}
\caption{(Color online) Different stackings of TMDs bilayers. Voilet and green balls are transition metal and chalcogen atoms. }
\label{stacking}
\end{figure}

While we have examined the onsite energies in going from monolayer to bilayer for the 2H stacking,
there are other stackings possible. In order to examine whether our conclusions were general enough,
we considered bilayers with AA, A$'$B, AB$'$ and AB stacking which are shown in Fig. ~\ref{stacking} \cite{cesare}.
As the environment for each atom in the monolayer as well as the bilayer are different  in each case, we expect
changes in the Madelung potential.
In each case the $ab-initio$ band structure was fit to a
tight binding model with Mo $d$ and Se $p$ states in the basis. The extracted onsite energies are given in
Table~\ref{mose3_all_stack}. These onsite energies for different stacking of layers in MoSe$_2$ bilayer
are similar to what we found in the case of 2H stacking (Table~\ref{onsite}).
Hence the changed environment for different stackings have very little effect on the onsite energies,
and electronic structure changes emerge from the differing interlayer interactions. This is consistent
with the approach adopted in the literature to examine the electronic structure of twisted bilayers ~\cite{twist_gh}. 
  Some of the previous studies on twisted bilayer graphene and silicene reported changes in local 
density of states in different stacking areas\cite{cite1,cite2}.

\begin{figure}[ht]
\centering
\includegraphics[height=16.0cm,width=7.0cm,angle=-270.0]{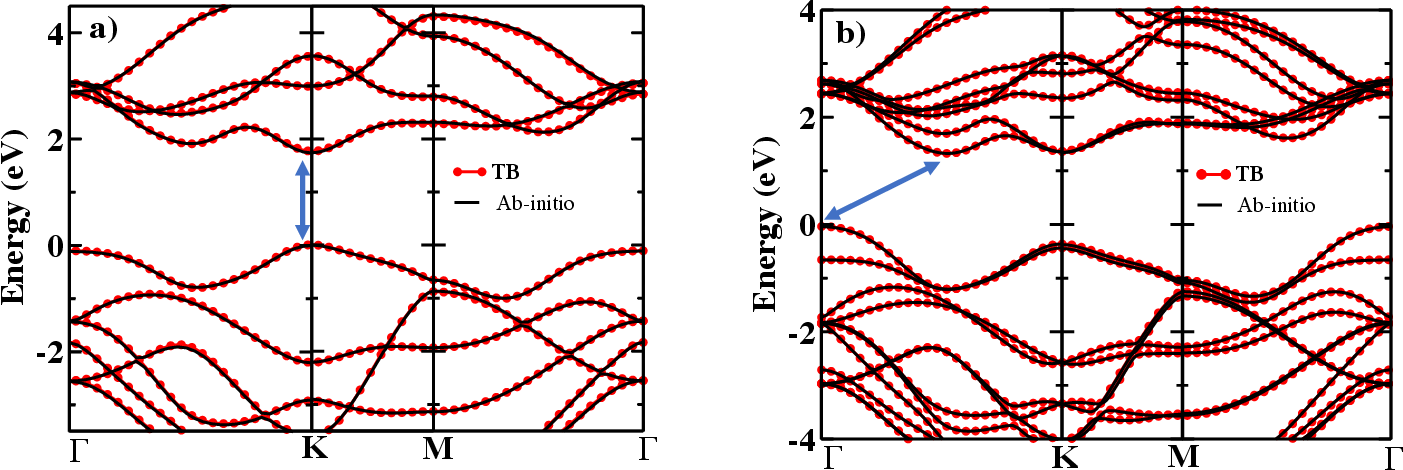}
\caption{ $ab-initio$ (solid line) and tight binding band structure (red circles) for (a) monolayer and (b) bilayer MoS$_2$. Zero of energy corresponds
to the valence band maximum. Arrows represent the VBM and CBM.}
\label{fig:7}
\end{figure}

The evolution of the electronic structure with number of layers that we find
here is not specific to MoSe$_2$ alone. We find similar changes when we
examine the electronic structure as a function of the number of layers
for MoS$_2$ also.
In Fig. \ref{fig:7} we have plotted the $ab-initio$ band
structure for the monolayer in panel (a) and for the bilayer in
panel (b). The tight-binding band structure has been superposed in each case
and we have a good description of the $ab-initio$ band structure.
A comparison of the extracted onsite energies is given in Table~\ref{onsitemos2}. These extracted values of the onsite energies are in agreement
with the values available in the literature \cite{prb_tb}.

\begin{figure}[ht]
\centering
\includegraphics[height=9.0cm,width=7.0cm,angle=-270.0]{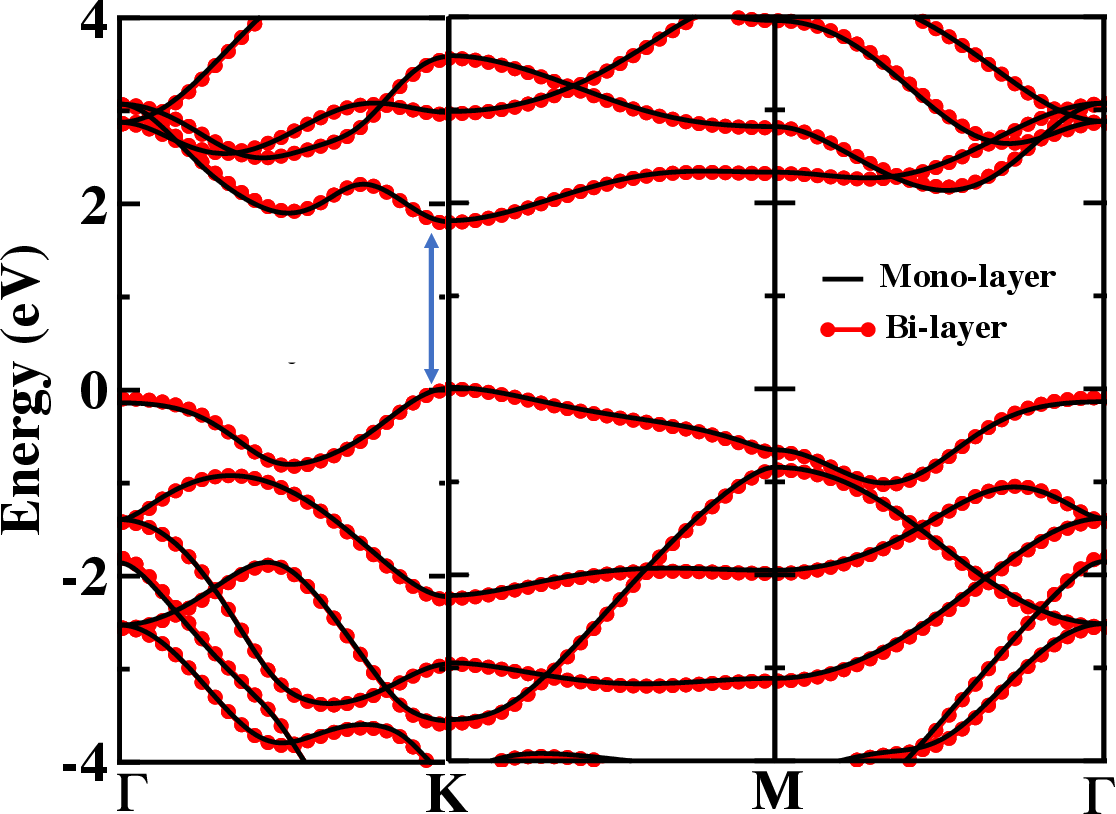}
\caption{The $ab-initio$ band structure (solid line) for monolayer MoS$_2$ compared with the tight binding band structure (red circles)
of bilayer with interlayer interactions switched off. The zero of energy is the valence band maximum. Arrows represent the VBM and CBM.}
  \label{fig:8}
\end{figure}

\begin{figure}[ht]
\centering
\includegraphics[height=16.0cm,width=11.0cm,angle=-270.0]{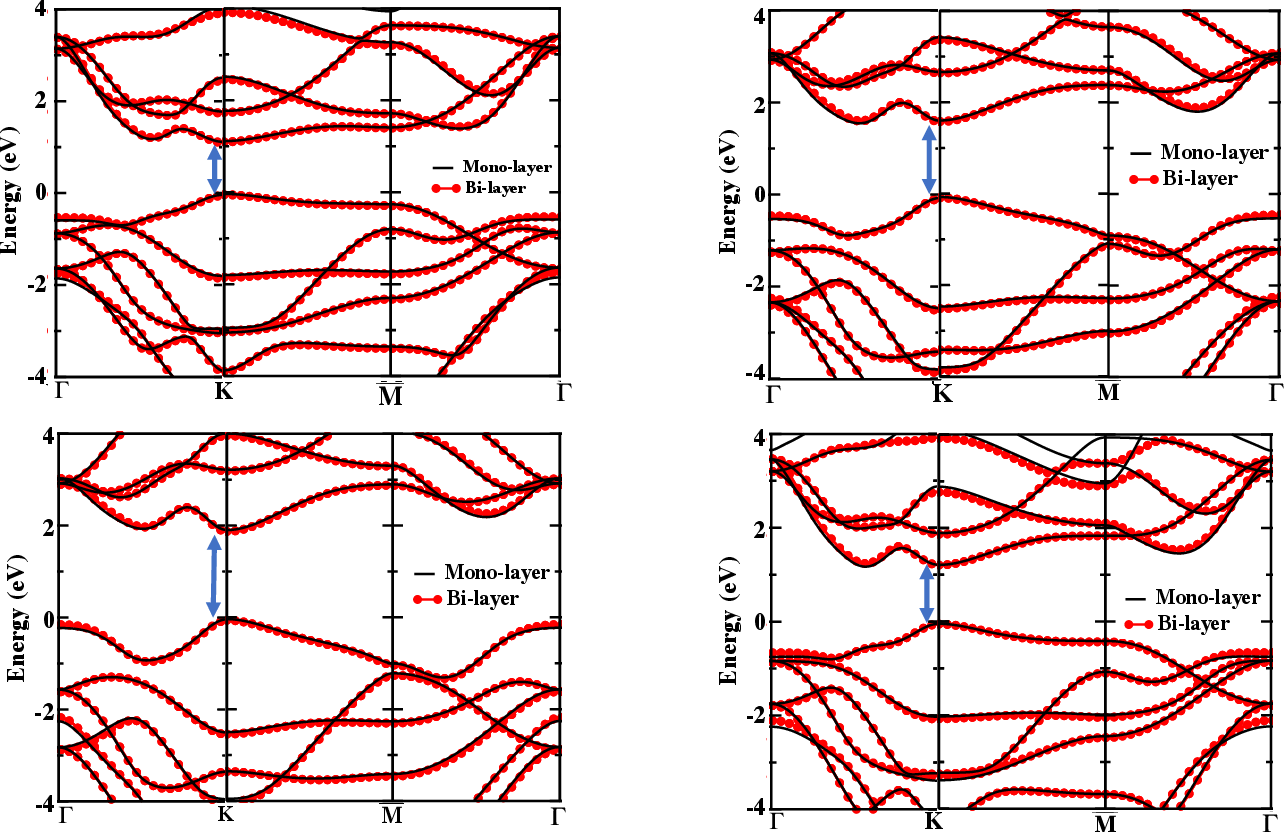}
\caption{(Color online) The $ab-initio$ band structure (solid line) for monolayer compared with the tight binding band structure (red circles)
of bilayer with interlayer interactions switched off, for (a) MoTe$_2$, (b) WS$_2$, (c) WSe$_2$ and (d) WTe$_2$. The zero of energy is the valence band maximum. 
Arrows represent the VBM and CBM.}
\label{fig:9}
\end{figure}

Here again, we find that the energies for the monolayer and bilayer are similar, as we found
earlier.
When we considered the tight binding
Hamiltonian for the bilayer and switched off interlayer interactions,
we recovered the monolayer band structure. The comparison between the tight binding
Hamiltonian results  with interlayer interactions switched off and the $ab-initio$ band structure for the
monolayer is given in Fig. \ref{fig:8}.
In order to demonstrate that the conclusions obtained from our analysis of MoSe$_2$ and MoS$_2$ are general, we have considered a monolayer as well as bilayers of MoTe$_2$,
WS$_2$, WSe$_2$ and WTe$_2$. The $ab-initio$ band dispersions calculated for the monolayer of each of these systems along various symmetry directions is shown in
Fig.\ref{fig:9}.
The  $ab-initio$ band structure for the bilayer was mapped onto a tight binding model with maximally localized Wannier functions for the radial part. The ensuing band structure
for the bilayers, with interlayer interactions switched off, has been superposed on the monolayer band structure in Fig.\ref{fig:9}. The two are found to be almost identical
for each of the systems shown here, indicating that the differences in the electronic structure between monolayer and bilayer arise due to interlayer interactions alone.

\section{Conclusions}
In conclusion, we have examined the evolution in the electronic structure of transition metal dichalcogenides as a function of layers. The changes in the structure that one
finds have been
discussed in terms of a combination of interlayer hopping interactions as well as Madelung potential effects.
In each case, a mapping of the $ab-initio$ electronic structure
onto a tight binding model with transition metal $d$ and anion $p$ states
in the basis allows us to quantify the role of each of these effects. Even in these layered materials which are referred to as van der Waals heterostructures, we find that
interlayer hopping interactions play the primary role in bringing about changes in the electronic structure as a function of thickness.
Expected Madelung potential variations on the other hand, we find, play no role in the observed changes in electronic structure.
These ideas are valid across the MX$_2$ where M = (Mo, W) and X = (S, Se, Te). While most results discussed in the manuscript correspond
to the 2H stacking, we show that considering other types of stackings does not change the conclusions.

\section{Methodology}
The electronic structure calculations of monolayer and bilayer MX$_2$ (M=Mo,W X=S,Se,Te)
have been carried out within a plane wave
implementation of density functional theory (DFT) within
VASP \cite{Kresse} (Vienna $ab-initio$ Simulation Package). We have taken the 2H 
stacking of the bilayers in each case as it is found to be the most stable
structure~\cite{cesare}. While, the lattice constants are kept at the experimental values
 for MX$_2$ series (M=Mo,W X=S,Se,Te )\cite{mos2_se2_exp, wse2_exp, ws2_se2_exp, wte2_mote2} which are listed in Table~\ref{lc_encut} ,  all the atoms are allowed to relax through a
total energy minimization that is guided by the calculated atomic forces.
A vacuum of 20 \AA{} is used along $z$ direction to minimize
the interaction among the periodic images for the monolayer. Projected augmented wave \cite{PAW,PAWpotentials1}
potentials are used to solve the electronic structure self-consistently using a k-points mesh of
12$\times$12 $\times$1. Cutoff energies for the plane wave basis states in material series MX$_2$ (M=Mo,W X=S,Se,Te) 
are also listed in Table~\ref{lc_encut} .
Perdew-Burke-Ernzerhof (PBE) \cite{PBE} approximation was used for the exchange-correlation functional.
The weak van
der Waals interaction between the layers has an effect in the determination
of the interlayer distances.
A dispersion correction based on Grimme's DFT-D2 method \cite{Grimme} is used
on top of the PBE potentials.

In order to quantify the results, we setup the following tight binding model
with the transition metal $d$ and anion $p$  states in the basis. 
\begin{align} \nonumber
   H & = \sum_{i,l} \epsilon_d d^{\dagger}_{il} d_{il} +   \sum_{i,l} \epsilon_p p^{\dagger}_{il} p_{il} \\ \nonumber
    & - \sum_{i,j,l_{1},l_{2}} \left( t^{l_{1}l_{2}}_{i,j,pd} d^{\dagger}_{il_{1}} p_{jl_{2}} + {\bf H.c.} \right) \\ \nonumber
        &- \sum_{i,j,l_{1},l_{2}} \left( t^{l_{1}l_{2}}_{i,j,dd} d^{\dagger}_{il_{1}} d_{jl_{2}} + {\bf H.c.} \right) \\ \nonumber
    &- \sum_{i,j,l_{1},l_{2}} \left( t^{l_{1}l_{2}}_{i,j,pp} p^{\dagger}_{il_{1}} p_{jl_{2}} + {\bf H.c.} \right) \\ \nonumber
\end{align}
where $d^{\dagger}_{il} \left(d_{il} \right)$ creates (annihilates) an electron in the $l$th $d$ orbital on transition metal 
site in the $i$th unit cell while $p^{\dagger}_{il} \left( p_{il} \right)$ creates (annihilates) an electron in the 
$l$th $p$ orbital on oxygen atom in the $i$th unit cell.
In this model,
the maximally localized Wannier
functions \cite{wannier90} are used for the radial parts of the basis functions.
Technically, the degree of localization and the symmetry
of these Wannier functions can be controlled in the
projection procedure. All on-site energies and hopping interaction strengths in this case are
determined from the interface of VASP to Wannier90 \cite{VASPtoWannier}. 
Once a full tight-binding Hamiltonian is obtained for the bilayers, in order to switch off the 
interlayer interactions, we identify 
all the interlayer terms in $ t^{l_{1}l_{2}}_{i,j,pd} d^{\dagger}_{il_{1}} p_{jl_{2}} $ and 
$  t^{l_{1}l_{2}}_{i,j,pp} p^{\dagger}_{il_{1}} p_{jl_{2}}$ and put corresponding $t$'s to zero.
Apart from the 2H stacking, for MoSe$_2$, we also explored
other stacking geometries AA, A$'$B, AB$'$ and AB  shown in Fig. \ref{stacking}\cite{cesare} to explore
the renormalization of the onsite energies due to differing
Madelung potentials. 

\section{Author Information}
S. K. Pandey and Ruma Das contributed equally to this work. Queries should be addressed to priya.mahadevan@gmail.com.

\vskip 0.5cm

\vskip 0.5cm

% \begin{figure}[ht]
% \centering
% \includegraphics[height=9.5cm,width=10.0cm]{Fig6.eps}
% \caption{ (a) The $ab-initio$ band structure for monolayer (dashed lines) and bilayer (solid lines) for MoSe$_2$ along various symmetry directions.
% The bands have been aligned at the valence band maximum at K point. Three sets of bands have been identified and are labelled A, B and C. The bands at B are
% zoomed in panel (b). The corresponding charge densities for A, B and C are shown in panels (c)-(e).}
%   \label{fig:6}
% \end{figure}

\vskip 0.5cm

\begin{table}
\caption{Lattice constants and plane wave energy cut-offs considered in our $ab-initio$ calculations for MX$_2$ series 
(M=Mo,W and X=S, Se and Te).} \vskip 0.5cm
\centering
\begin{tabular}{c c c}
\hline\hline
  Material & Lattice constant (\AA{}) & Cut-off energy (eV) \\ [1ex]
\hline
MoS$_2$ & 3.150 & 350 \\
MoSe$_2$ & 3.289 & 280 \\
MoTe$_2$ & 3.519 & 280  \\
WS$_2$  & 3.153& 350 \\
WSe$_2$& 3.282 & 280  \\
WTe$_2$ & 3.491 & 280  \\[1ex]
\hline\hline
\label{lc_encut}
\end{tabular}
\end{table}

\begin{table}[ht]
\caption{Onsite energies obtained from tight binding mapping of the $ab-initio$
band structure. A basis consisting of Mo \textit{d} and Se \textit{p} states has been considered for monolayer and bilayer MoSe$_2$.
The respective Se $p_x$ is taken as reference for the energies given in eV. } \vskip 0.5cm

\centering
\begin{tabular}{c c c}
\hline\hline
 Orbitals & Monolayer MoSe$_2$ & Bilayer MoSe$_2$ \\ [1ex]
\hline
Se-$E_{p_x}$ & 0.00 & +0.01 \\
Se-$E_{p_y}$ & 0.00 & 0.0 \\
Se-$E_{p_z}$ & -0.33 & -0.38 \\
Mo-$E_{d_{xy}}$ & +1.46 & +1.51 \\
Mo-$E_{d_{yz}}$ & +2.28 & +2.31 \\
Mo-$E_{d_{zx}}$ & +2.28 & +2.31 \\
Mo-$E_{d_{x^2-y^2}}$ & +1.46 & +1.52 \\
Mo-$E_{d_{z^2}}$ & +1.20 & +1.27 \\[1ex]
\hline\hline
\label{onsite}
\end{tabular}
\end{table}

\begin{table}[!ht]
\caption{Onsite energies obtained from tight binding mapping of the $ab-initio$
band structure for different stackings of layers in  MoSe$_2$ bilayer. A basis consisting of Mo \textit{d} and Se \textit{p} states has been considered. The respective Se $p_x$ is
taken as reference for the energies given in eV.} \vskip 0.5cm
\centering
\begin{tabular}{c c c c c c c}
\hline\hline
 Orbitals & AA & A$'$B & AB$'$ & AB \\ [1ex]
\hline
Se-$E_{p_x}$ & 0.0 & 0.0 & 0.0 & 0.0 \\
Se-$E_{p_y}$ & 0.0 & 0.0  & 0.0 &  0.0 \\
Se-$E_{p_z}$ & -0.35 & -0.36 & -0.35 & -0.35 \\
Mo-$E_{d_{xy}}$ & +1.52 & +1.53  & +1.50 & +1.51 \\
Mo-$E_{d_{yz}}$ & +2.30 & +2.31  & 2.29 & +2.30 \\
Mo-$E_{d_{zx}}$ & +2.30 & +2.31  & 2.29 & +2.30 \\
Mo-$E_{d_{x^2-y^2}}$ & +1.52 & +1.53  & +1.50 & +1.51 \\
Mo-$E_{d_{z^2}}$ & +1.28 & +1.28  & +1.25 & +1.26 \\[1ex]
\hline\hline
\label{mose3_all_stack}
\end{tabular}
\end{table}

\begin{table}
\caption{Onsite energies obtained from tight binding mapping of the $ab-initio$
band structure. A basis consisting of Mo \textit{d} and S \textit{p} states has been considered for monolayer and bilayer MoS$_2$ . 
The respective S $p_x$ is taken as reference for the energies given in eV. In the fourth column, onsite energies from reference\cite{prb_tb} has been listed.}\vskip 0.5cm
\centering
\begin{tabular}{c c c c}
\hline\hline
 Orbitals & Monolayer MoS$_2$ & Bilayer MoS$_2$ & BL MoSe$_2$\cite{prb_tb} \\ [1ex]
\hline
S-$E_{p_x}$ & 0.00 & 0.0 & 0.0\\
S-$E_{p_y}$ & 0.00 & 0.0 & 0.0 \\
S-$E_{p_z}$ & -0.28 & -0.25 & -0.52 \\
Mo-$E_{d_{xy}}$ & +1.66 & +1.69 & +1.39\\
Mo-$E_{d_{yz}}$ & +2.68 & +2.69 & +2.36\\
Mo-$E_{d_{zx}}$ & +2.68 & +2.69 & +2.36\\
Mo-$E_{d_{x^2-y^2}}$ & +1.65 & +1.68 & +1.39\\
Mo-$E_{d_{z^2}}$ & +1.46 & +1.47 & +1.15\\[1ex]
\hline\hline
\label{onsitemos2}
\end{tabular}
\end{table}

\pagebreak

\newpage
\bibliography{MoX2_acs}

\end{document}